# Controllable switch of dissolution kinetics mechanism in drug microcapsules

Ushkov A.[1,*], Machnev A.[1], Ginzburg P.[1]

[1]School of Electrical Engineering, Tel Aviv University, Tel Aviv 69978, Israel

**Abstract**

Controllable continuous release of functional materials from capsules is one of the unmet functions of theragnosis particles; on this way, understanding cargo-fluid interactions *in vitro* is an essential milestone. We develop a flexible platform to investigate single particle-fluid interactions utilizing a glass micropipette as a highly localized flow source around an optically trapped particle. In proof-of-concept experiments this microparticle is sensitive to local microflows distribution, thus serving as a probe. The very same flows are capable of the particle rotating (i.e., vaterite drug cargo) at frequencies dependent on the mutual particle-pipette position. The platform flexibility comes from different interactions of a tweezer (optical forces) and a pipette (mechanical/hydrodynamical) with micro-particle, which makes this arrangement an ideal micro-tool. We studied the vaterite dissolution kinetics and demonstrated it can be controlled on demand, providing a wide cargo release dynamic rate. Our results promote using inorganic mesoporous nanoparticles as a nanomedicine platform.

*corresponding author e-mail: andreiushkov@tauex.tau.ac.il

**Main**

The vaterite polymorph of calcium carbonate (CaCO3) has a wide scope of potential applications in biology and medicine due to its outstanding combination of chemical, morphological, and physical properties [1] [2]. Being a fully biocompatible, FDA-approved material, it possesses a sponge-like morphology that effectively entraps a variety of functional materials. As the least stable phase of $CaCO_3$, vaterite spherulites require mild conditions for phase transforms (i.e., to calcite) or dissolution, thus facilitating their use as smart drug vehicles for targeted drug delivery [3]. In addition, vaterite synthesis is a cheap, simple yet flexible process, yielding crystals of different shapes in dependence on chemicals used [4]. As a cargo release from vaterite particles is associated with structural transformations, an understanding of particle-fluid interactions is required for the future advancement of this material platform toward targeted drug delivery needs. Furthermore, propulsion and diffusion of capsules are also dependent on interaction with a fluid environment, further motivating to perform detailed studies with full control over parameters. A combination of several microtools can allow performing those types of studies.

Pre-pulled glass micropipettes are must-have tools for diverse scientific needs, from biological engineering [5][6] and microinjection [7][8] to electrophysiology [9][10][11]. In a broad sense, these instruments are beneficial for their ability to build a bridge between micrometer-sized objects (bacteria, microparticles, or long DNA chains) and macroscopic controlling tools (stepper motors for mechanical manipulations, an electronic amplifier for ionic currents registration, microsyringe pump for drug injections).

Micropipette-assisted particle manipulation is a relatively novel yet prospective field with applications in targeted drug delivery, non-contact sorting of precious cells in microfluidic channels, and many others. In [12] conical glass micropipettes were used to measure mechanical properties of soft polymer gel microparticles in a recently developed microaspiration technique. In order to obtain the time-resolved information about the mass of selected particles and unicellular organisms, authors of [13] have constructed the micropipette-based mechanical resonator. Since most of the microparticles being studied are considered in a colloidal phase, microfluidic effects play an important role in their behavior and even may, for example, lead to hydrodynamic trapping [14]. Micropipettes with mounted piezoelectric elements are capable of generating liquid vortices in the vicinity of the particle, pulling it towards the pipette, and transporting it to a new position [15]. The internal hollow channel of micropipettes brings an additional degree of freedom in microfluidics, as it allows releasing a limited dose of liquid or supporting a pressure-driven flow from the orifice. Thus, the researchers demonstrated real-time control over single nanoparticle trajectories by balancing the pressure and electric field forces [16]. The tracking of particle trajectories in a liquid flow inside the glass capillary allowed the construction of the subnanoliter-precision piezoelectric pipette [17].

The progress in calcium carbonate drug capsules moved from the study of macroscopic volumes [18] [19] with millions of particles to a microfluidic reactors research [20] [21], which gives a better real-time insight into the precipitation/dissolution kinetics in the presence of local hydrodynamics. The question of individual drug cargo-fluid interaction becomes even more important in the context of natural microfluidic systems - obstructed blood vessels [22] [23]. Consequently, a growing need exists for a simple yet flexible platform for real-time testing of drug carrier-fluid interactions at a single particle level. We address this issue by integrating two common microtools - optical tweezer and pre-pulled glass micropipette - in one experimental setup. We consider interactions between a of the system under study (particles 4-5 µm in diameter outside the pipette in the vicinity of its orifice microcapsule and a local hydrodynamics generated by open-ended micropipette. Characteristic size ~1 micron in diameter) corresponds to typical objects considered in microfluidic systems, bacteria, and colloids. Although such a system is comparatively known [24][25][26][27], a pipette there is usually utilized as a mechanical holder with a precise, but motionless, positioning. Though smaller vaterite particles are considered in drug delivery applications, investigating interactions with larger and more controllable samples provides guidelines for nanoscale dynamics. Here we mount it on a motorized device to turn it into a fully functioning 3D manipulator. Since it interacts with microobjects (mechanically or hydrodynamically) differently than an optical tweezer (optical forces), these two micromanipulation tools don't interfere and ideally complement each other. We demonstrate that in liquid environments the micropipette can act as a delicate tool for micromechanical manipulations, capable of applying on objects under study the hydrodynamic forces of the same magnitude as those in optical tweezers. Furthermore, a fully automated pipette motion allowed the flow pattern visualization. The second observed effect is that the very same liquid flows imply a torque to the trapped particles for certain particle-pipette orientations. The rotation frequency was measured via a photodetector scheme for the case of birefringent vaterite spherulites. Experimental data is in qualitative correspondence with finite-element simulations.

The described integrated setup may serve as a flexible platform to study microparticles-fluid interactions on a single particle level. Although we observed vaterite spherulite rotations only, the method can be used for a wide range of optically trappable micro-objects with spherical or non-spherical geometries, smooth and rough surface morphology. Micropipette-driven flows allow rotating particles irrespectively to their optical anisotropy and trapping beam polarizations, being a simplified analog of optically actuated microtools [28] [29] [30].

The manuscript is organized as follows – the set of optical tweezer-assisted microfluidic tools is established first, tested and the validity of the methodology is verified with finite element numerical analysis. The dissolution dynamics of drug carriers is then revealed and supported by a new theoretical model, which allows for predicting the drug release rates. Experiments show that the release rate can be controlled in a range of minutes to days, making the vaterite-dissolvable platform extremely relevant to drug delivery applications.

**Experimental Layout**

The experimental bench is sketched in Fig.1. An optical tweezer setup, mounted on an inverted microscope architecture, was implemented [31]. A Petry dish was used as a liquid cell; a hole in the bottom of the dish was drilled and covered with a microscope cover glass 0.17 mm in thickness in order to use the optical tweezer in an inverted microscope configuration.

For the glass micropipette pulling the Narishige PC-10 Puller in a double pull mode was used, the micropipette average orifice was adjusted to ~1 μm in diameter. After pulling, the glass pipette was accurately bent near the tip as described in [32] in order to bring the pipette towards microparticles as horizontally as possible, i.e. to keep the bended part of the tip in the image plane. The pipette was mounted on the Scientifica PatchStar micromanipulator to achieve full control over the pipette movement in three dimensions. To facilitate and speed up the experiments a custom program in Matlab was written, which sends ASCII commands to the micromanipulator via the serial COM port. A liquid flow from the pipette orifice is induced by applying pressure from a hydrostatic column. Manipulations with particles were observed and recorded by a video camera. Quadrant photodetector additionally recorded the periodical scattering signal from rotating vaterite particles.

For the experiments two types of microparticles were used: commercial Sigma-Aldrich silica beads 5 micron in diameter, and synthesized vaterite spherulites 5 micron in diameter. All experimental manipulations were performed in ethanol in cases the stability of the vaterite phase is required, otherwise particles might undergo a calcite phase with time [33]. The dissolution dynamics was performed in deionized water. Commercial silica beads are covered with long surfactant molecules to prevent particle agglomeration, however, it might affect the mechanical behavior of optically trapped particles in a liquid flow. Thus, for cleaning purposes, before the experiments, silica beads passed three iterations of sonication, centrifugation, and the subsequent renewal of the ethanol solution.

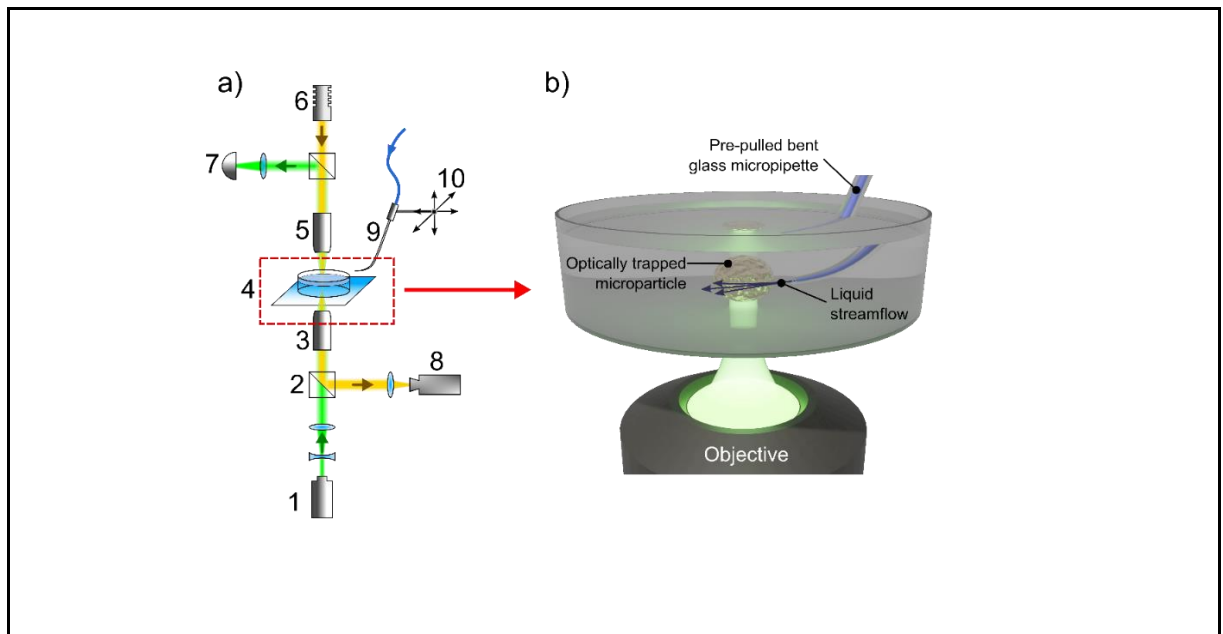

Figure 1. (a) A scheme of the optical tweezer and micropipette setup: 1 - monochromatic laser source (980 nm wavelength); 2 - beam splitters; 3 - 100x objective; 4 - sample holder and liquid cell; 5 - 10x objective; 6 - white light source; 7 - quadrant photodetector; 8 - video camera; 9 - pre-pulled glass micropipette; 10 - pipette holder mounted to the micromanipulator. (b) Principle of micropipette-powered dynamics of optically trapped microparticles: liquid flows from the pipette orifice, affecting the particle motion, whereas the optical trap holds the object.

**Liquid flow measurements**

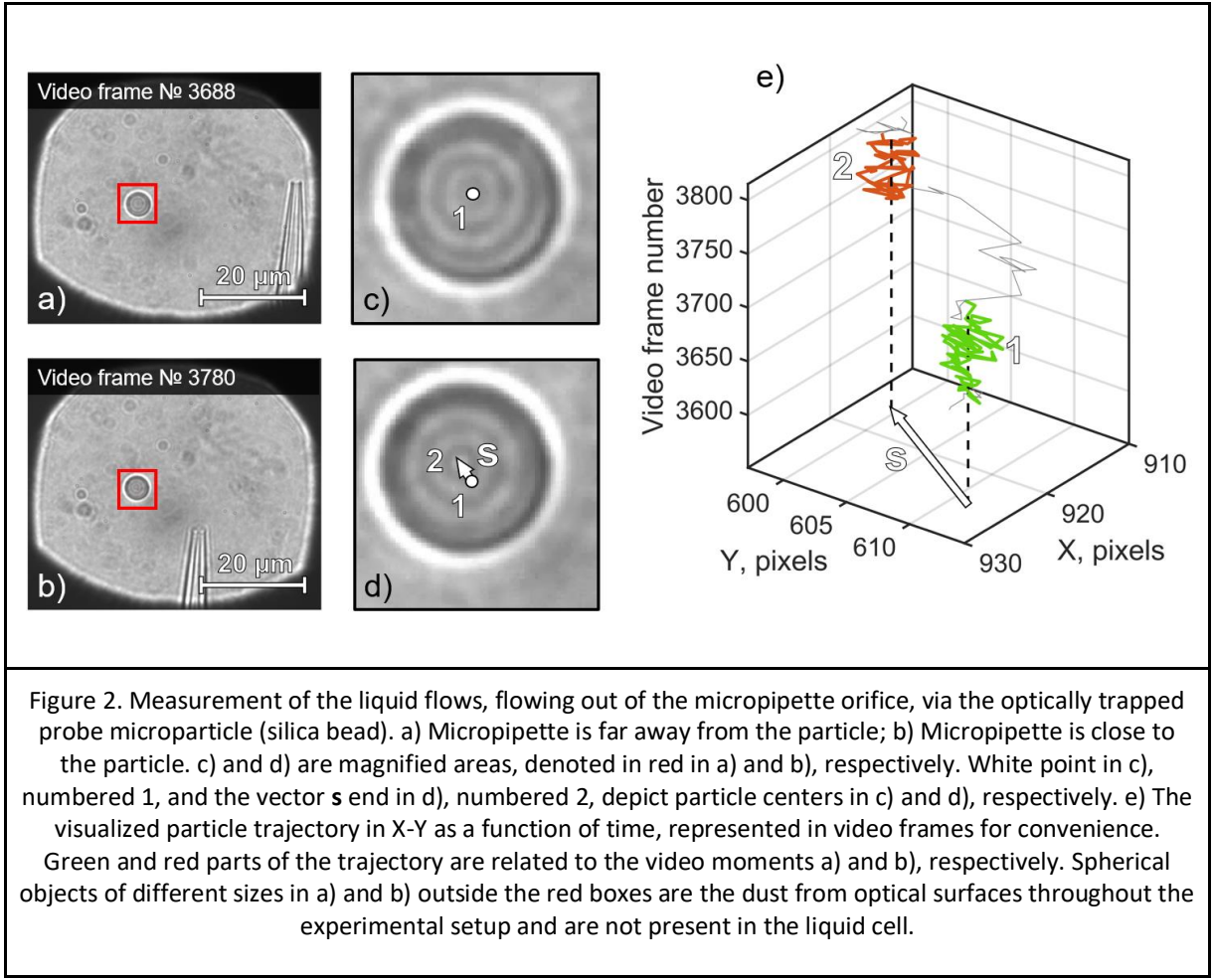

Figure 2. Measurement of the liquid flows, flowing out of the micropipette orifice, via the optically trapped probe microparticle (silica bead). a) Micropipette is far away from the particle; b) Micropipette is close to the particle. c) and d) are magnified areas, denoted in red in a) and b), respectively. White point in c), numbered 1, and the vector **s** end in d), numbered 2, depict particle centers in c) and d), respectively. e) The visualized particle trajectory in X-Y as a function of time, represented in video frames for convenience. Green and red parts of the trajectory are related to the video moments a) and b), respectively. Spherical objects of different sizes in a) and b) outside the red boxes are the dust from optical surfaces throughout the experimental setup and are not present in the liquid cell.

The first effect measured in our study is the deviation of the optically trapped particle from its equilibrium position due to the pressure-driven liquid flows from the micropipette orifice. In the initial set of studies, we take a smooth silica particle as a test object to assess the setup capabilities. Figure 2 shows the measurement process. The equilibrium position of the optically trapped particle (position 1), with the micropipette being far away from it, is shown in Figs.2a,c. Then the micropipette was moved to the specified position in the vicinity of the particle (Fig. 2b, position 2), and a particle shift **s** was observed (Fig.2d). The Brownian motion introduces high-frequency fluctuations of the particle position; the Mean Squared Displacement (MSD) of the optically trapped particle in an overdamped regime is known to be [31]:

$$MSD_x(\tau) \equiv \overline{[x(t+\tau) - x(t)]^2} = 2\frac{k_B T}{\kappa}[1 - exp(-|\tau|/\tau_{to})], \qquad (1)$$

where $\kappa$ is the trap stiffness, $\gamma = 6\pi\eta a$ is the friction coefficient, determined by Stoke's law for spherical particles, and $\tau_{to} = \gamma/\kappa$ is the characteristic time, when the particle "feels" the trap restoring force. Taking into account the actual silica bead diameter $2a = 5$ μm, ethanol dynamic viscosity $\eta \approx 1.144$ mPa*s, and $\kappa \approx 10$ pN/μm, we estimate $\tau_{to} \approx 5$ ms. The particle center coordinates $x(t_i), i = 1, \ldots, N$ were measured during $t_m = 1$ s in both positions 1 and 2 at 40 fps. The relationship $t_m \gg \tau_{to}$ guarantees that the particle has enough time to oscillate in the optical trap around its equilibrium position, defined as a mean $x_{eq} = \frac{1}{N}\sum_{i=1}^{N} x(t_i)$. The particle shift between positions 1 and 2 due to the liquid flow is $\mathbf{s} = x_{eq,2} - x_{eq,1}$. Once the particle shift was obtained, it allows one to estimate the liquid flow direction and velocity for the current particle-pipette relative position using the Stoke's law: $\mathbf{u} = \mathbf{s}/\gamma$.

The described velocimetry method utilizes the probe particle with diameter of $2a = 5$ µm, which refers to the upper size limit of standard flow tracers (20-3000 nm in diameter) for Micro-Particle Image Velocimetry (µ-PIV) [34]. In µ-PIV smaller particles with $2a \lesssim 1$ µm usually give better spatial resolution and less affect the measured flows. However, our choice of characteristic particle sizes is dictated by two reasons. Firstly, due to the diffraction limitations particles $2a \lesssim 1$ µm are difficult to observe in standard optical microscopes, thus more complicated setups, e.g., with fluorescent materials and laser pumping are required. Secondly, the optically trapped particle should possess a certain mobility so we could detect the liquid flow via the particle shift. As it was studied experimentally [35][36] and theoretically [37][38], optically trapped silica particles with diameters $2a \gtrsim 1$ µm, embedded in water or ethanol, follow the long-wave (ray optics) approximation regime with a trap stiffness $\kappa \propto 1/a$. Thus, the increase of particle size allows one to controllably reduce the trap stiffness until the particle shift is detected reliably in a certain microscope setup.

The automatic 2D sweep of micropipette positions in the vicinity of the trapped particle generates 2D flow fields near the pipette orifice. Figure 3 demonstrates these fields for two hydrostatic pressures applied to the pipette: ~2200 Pa (Fig.3a) and ~2900 Pa (Fig.3b).

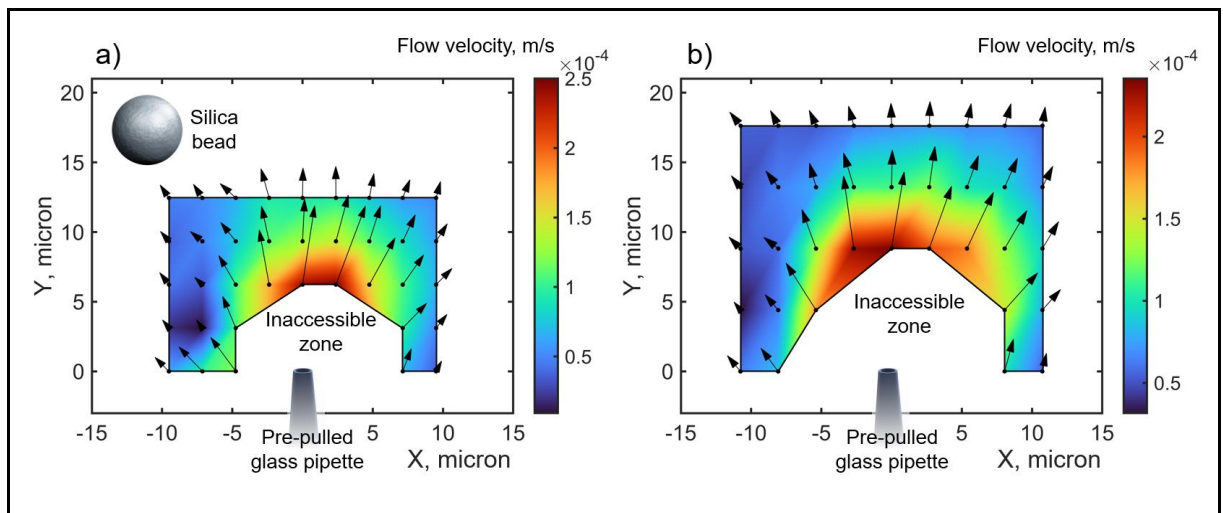

Figure 3. Flow fields near the micropipette orifice, experimentally measured via the silica particle shift in the optical trap as shown in Fig.2, for hydrostatic pressure applied to the pipette: a) ~2200 Pa, b) ~2900 Pa. Colors depict the flow velocity value, black arrows - velocity value and direction. Inaccessible zones in a) and b) appear because the particle cannot physically go through the pipette, and due to the liquid flows, which are strong enough in this region to sweep the particle away from the optical trap. The size of a silica bead in a) is proportional to the experimental region of interest.

**Flow-driven micro-object rotations**

After revealing the setup capabilities in measuring microfluidic interactions, vaterite particles were assessed. Furthermore, those also allow testing other types of mechanical motion. Another type of micropipette-powered dynamics are particle rotations. We study the rotation of birefringent vaterite spherulites with diameter of $2a = 5$ µm, which are the best match for two reasons. Firstly, their total torque in a liquid flow is larger than that for smooth spherical particles. The reason is that in addition to the viscous stress tensor term, it has a contribution proportional to the liquid pressure [39] due to the natural surface roughness. Secondly, the rotating vaterite spherulite periodically "flickers" due to its birefringence, which allows one to easily measure the rotation frequency by a photodetector [40] (see Fig.1a). Though smaller vaterite particles are usually utilized as drug carriers, we consider larger samples to attain more controllable experimental conditions, which however provide guidelines for nanoscale dynamics. Micropipette-driven flows allow rotating particles irrespectively to their optical anisotropy and trapping beam polarizations, being a simplified analog of optically actuated microtools. In the current report, however, we restrict ourselves to the vertical rotation axis regime as a simplest case for experimental measurements.

General considerations suggest that a maximum rotating frequency is achieved for the particle located on the sides of the pipette, where a maximum difference of flow velocities around the particle happens. This intuitive assumption is confirmed by finite-element simulations, see Fig. 4. A simplified 2D model was considered: a solid particle iteratively sweeps a rectangular mesh of positions around the pipette. A simulation box of the size 40 by 30 $\mu$m has been used with an outlet boundary conditions (p = 0 Pa) at the edges in order to avoid affecting by the edges of the box. At the particle-water interface we chose no slip boundary condition as well as to the water-pipette interface. This simulation serves for illustration purposes only, so only general geometrical parameters - particle and pipette orifice sizes - have been taken from experiments. The calculated torque was normalized to unity.

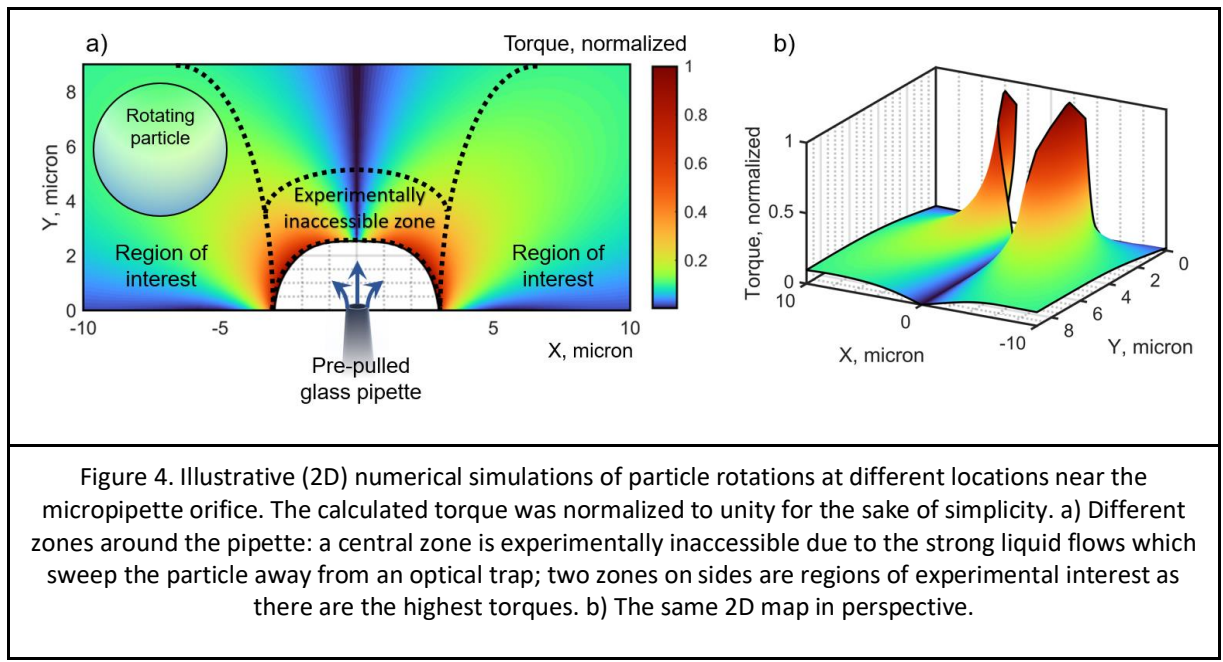

Figure 4. Illustrative (2D) numerical simulations of particle rotations at different locations near the micropipette orifice. The calculated torque was normalized to unity for the sake of simplicity. a) Different zones around the pipette: a central zone is experimentally inaccessible due to the strong liquid flows which sweep the particle away from an optical trap; two zones on sides are regions of experimental interest as there are the highest torques. b) The same 2D map in perspective.

The 2D map of particle rotations in Fig.4 shows that a region along a central vertical axis X=0 has only close-to-zero frequencies and is therefore out of experimental interest. In addition, it contains the experimentally inaccessible zone close to the pipette, where the liquid flow is strong enough to sweep the vaterite away from the optical trap. Consequently, there are two symmetric regions of interest for experiments, located on both sides of the pipette.

The process of experimental measurements is analogous to those presented in Figs.2a,b, with the difference that the particle rotational dynamics is measured via the photodetector, and not via the image processing, see Figs.5a-e. The automated process of the micropipette moving through the prescribed positions allows one to get 2D maps of rotation frequencies, see Figs.5f-g.

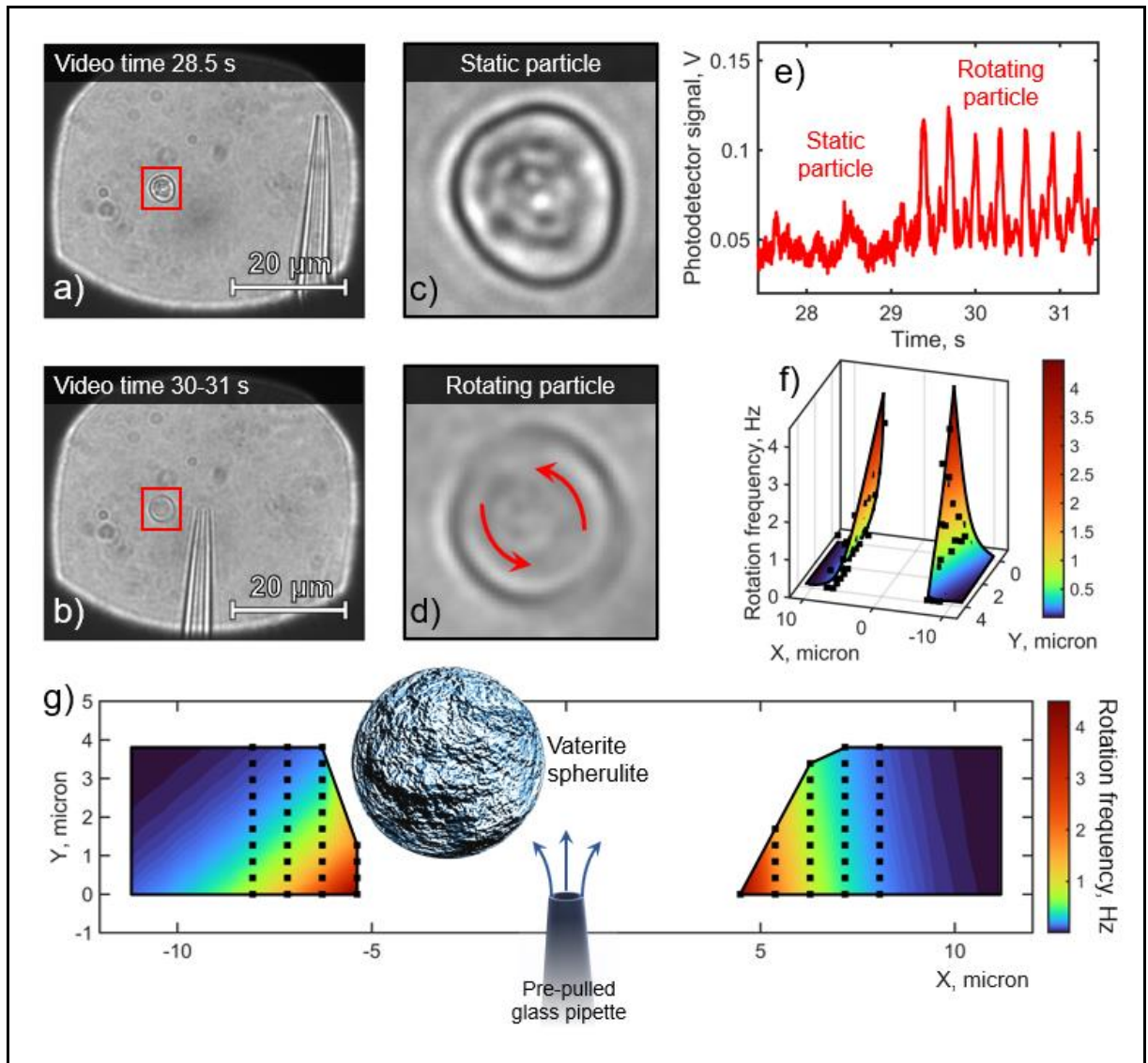

Figure 5. Measurement of the rotational dynamics of a vaterite microparticle, trapped in the optical tweezer. Rotations are caused by liquid flows, flowing out of the micropipette orifice. The hydrostatic pressure applied to the pipette is ~2900 Pa. a) Micropipette is far away from the particle; b) Micropipette is close to the particle. c) and d) are magnified areas, denoted in red in a) and b), respectively. The particle in c) is static, in d) is rotating. Image d) averages ~40 frames from the video, in order to demonstrate the particle smoothing due to rotations. e) Photodetector signal from a vaterite in static and rotating regimes. Spherical objects of different sizes in a) and b) outside the red boxes are the dust from optical surfaces throughout the experimental setup and are not present in the liquid cell. f)-g) Measured 2D distribution of rotation frequencies near the pipette orifice, located in coordinates (0,0). Black squares denote the experimental data. The vaterite spherulite in g) has a size proportional to experimental Region of Interest.

The experimental results are in qualitative agreement with simulations in Fig.4: the highest frequency is in regions closest to the pipette on both sides from it, and quickly drops down when moving away. Despite the big vaterite size (5 μm in diameter), the frequency is quite sensitive to the particle-pipette distance: it decreases by half, from 4 Hz to 2 Hz, when the pipette is moved 1 μm away (see the region (X,Y)=(5,0) μm in Fig.5g).

**Switching of the vaterite dissolution kinetics mechanism using a micropipette flow**

The vaterite spherulites dissolution kinetics in different liquid environments is crucially important for morphology investigations [41], phase stabilization [42], and controlled drug release [43]. More generally, nano/microparticle dissolution rates are a key characteristic for developing pharmaceutical and agricultural products [44], [45], and take a central place in various biodurability tests [46]. Typically, crystal phase growth and subsequent dissolution are usually studied in mother solution with high precipitate concentrations and strong magnetic stirring [47,48], therefore mainly the statistical information obtained from pH meters, optical spectrometers, and diffractometers are analyzed. In the case of calcium carbonate, the interaction dynamics is governed by the precipitation and transformation between three anhydrous polymorphs: vaterite, aragonite, and calcite [49].

The crystal kinetics studies may become much more controllable at a single-particle level [41][50][51][52], since it allows direct rate measurements for certain processes at interest. In this Section we demonstrate the possibilities of the micropipette-integrated optical tweezer in the domain of crystal kinetics.

An apparent feature of the setup is the possibility to optically trap a single particle and intentionally separate it from other crystallites by a long distance, either by lifting it up above the substrate, or by moving it along a surface towards regions with low particle concentration. If the trapped particle is ~10 diameters away from other crystals in a stagnant solution, the diffusion field around it can be approximated as a stationary with a central spherical ionic source of the same size as the trapped particle [53]. In this case, the steady-state ions concentration around the vaterite follows a law [48][53] (Fig.6b):

$$C(r) = \begin{cases} C_s, & 0 \leq r \leq R \\ \dfrac{R(C_s - C_{bulk})}{r} + C_{bulk}, & r \geq R \end{cases} \qquad (2)$$

where $C(r)$ is a species concentration at a distance $r$ from the particle center, $R$ is the particle radius, $C_s$ is a vaterite solubility, and $C_{bulk}$ is a species concentration far away from the particle. As the concentration at the vaterite/liquid interface is equal to the vaterite solubility, the particle dissolution process is limited by the transport of dissolved species away from the interface via diffusion, and is therefore called a diffusion-limited process. Applying Fick's first law for $C(r)$ in Eq.(1) and the integration of flows of matter around the sphere yields the particle size-dependent dissolution rate [53]:

$$\frac{dr}{dt} = \frac{DV_m[C_{bulk} - C_s]}{r} \cdot \ln(2) \equiv \frac{B}{r} \qquad (3)$$

where $D$ is an effective electrolyte ions diffusion constant, $V_m$ is a vaterite molar volume, and $\ln(2)$ takes into account the proximity of the particle to the substrate [54].

The diffusion-limited dissolution is typically valid for small particles up to ~10 microns [52][55]. In our setup, we verify the validity of this model via a direct real-time measurements $r(t)$ for a single optically trapped vaterite embedded in deionized water (Fig.6a, curve #1), and fitting it with a theoretical curve $r(t)$ obtained from Eq. (3) (Fig.6c). The theoretical value of $B$ from Eq.(3), calculated using tabulated values of $V_m$, $C_s$ and $D$, is $B^{theory}$=-0.0023 m$^2$/s (see Supplementary Materials), whereas the experimental data fitting yields $B^{exp}$=-0.0019 m$^2$/s. The proximity of these values confirms the hypothesis of the predominantly diffusion-controlled dissolution mechanism.

As it was mentioned above, small particles up to ~10 microns typically obey a diffusion-limited dissolution. This model is valid in stagnant solution, and even in the presence of convection and magnetic stirring [56], because small microparticles move with a speed of flow and their local diffusion field is still unperturbed. However, the behavior beyond the diffusion-limited model, when there is no immobile liquid shell around the particle and a dramatic fall in ion concentration occurs at the interface (Fig.6b, green curve), attracts a big interest [57][58]. In this case, a dissolution rate gives important information about the crystal surface, since it is governed by numerous mechanisms of nucleation and defects formation [55]. In contrast to the diffusion-controlled one, this dissolution regime is called surface-controlled. Although various surface-limited reactions may change the dissolution rate sufficiently, they nevertheless act as a constant factor in crystal kinetics with a constant dissolution rate independent from particle size [52][55], which is enough for our consideration.

In order to break the immobile layer around the optically trapped vaterite particle under study we employ a micropipette-generated flow of deionized water. Figure 6d demonstrates a dissolution behavior of such a particle with and without micropipette flow. In the first scenario (curve #3), the particle is being constantly washed by a water flow. The fluid velocity was measured to be $3.9*10^{-4}$ m/s, using a method with a trapped silica bead described above. In contrast to the diffusion-controlled regime (curve #1 in Fig.6a), this curve corresponds to the surface-controlled dissolution and demonstrates a constant dissolution rate. In a second scenario (curve #4), first ~20 seconds of experiment the particle was immersed into a pipette stream, the next ~40 seconds the pipette was moved far away, no external flow washed the trapped vaterite, and the last period (from t≈60 s to t≈100 s) the particle was washed in a pipette flow again. The curves show the clear difference in behavior without and with micropipette flow. It is worth noting that the dissolution rate for the flow-washed vaterite (curve #3 and partially curve #4 at t<20 s and t>60 s in Fig.6d) does not depend on the microparticle diameter. In contrast, the diffusion-controlled dissolution rate of the curve #4 at 20 s<t<60 s does depend on the particle size. Although the last statement is not clear from Fig.6d due to the short time of the diffusion-controlled regime in this particular experiment, it becomes evident from the curve #2 in Fig.6a. Two diffusion-controlled dissolution periods of time (that is, when the micropipette is far away from the particle) are denoted in gray on this curve, have different slopes and therefore two different dissolution rates. Consequently, two qualitatively different dissolution behaviors may be observed for the very same vaterite microparticle.

Summing up the current section, we have experimentally demonstrated a switching of the single vaterite dissolution between diffusion- and surface-controlled regimes, via the micropipette-generated liquid flow. Interestingly, as it was mentioned above, statistical dissolution kinetics of small (<10 micron) microparticles is usually considered diffusion-controlled even in the presence of convection [53]. Consequently, the considered micropipette-integrated optical tweezer setup allows observation of atypical crystal kinetic regimes. In addition, the setup unblocks a real-time changing of the microparticle size in a controllable manner, which might have applications in particle scattering and adjustable optical devices.

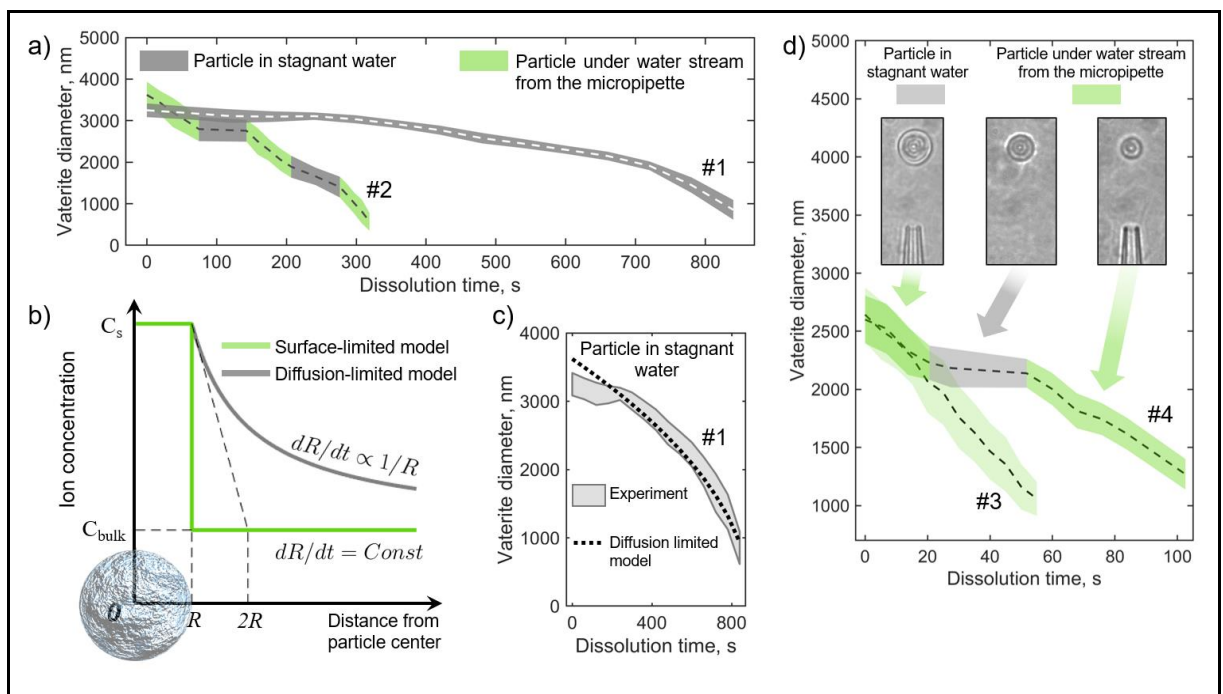

Figure 6. Dissolution behavior of optically trapped vaterite microparticles in deionized water with and without water flow from the micropipette. a) Dissolution curve #1 - in stagnant water (micropipette is far away); curve #2 - with alternating time periods with and without micropipette-generated flow. Color stripes along dashed lines denote measurement errors. b) Steady state dissolved species concentration around a single vaterite microparticle in surface- and diffusion-limited mode; c) Diffusion-limited dissolution of vaterite microparticle in stagnant water (gray stripe) and a theoretical curve (dashed line); d) Vaterite dissolution curve #3 - under a constant deionized water flow (~$4*10^{-4}$ m/s) from the micropipette orifice; curve #4 - with alternating time periods with and without micropipette-generated flow.

**Conclusions**

We propose a novel microfluidic approach to study both dynamics and kinetics of microparticles, using the pre-pulled glass micropipette as a source of liquid flow and an optically trapped microparticle as a probe. In proof-of-concept experiments, we reconstructed the liquid flows in the vicinity of the pipette orifice by considering the particle shift in an optical trap, studied the rotational dynamics and real-time dissolution kinetics of vaterite spherulite.

The proposed method of liquid velocity measurement may be effective for stationary flows in limited volumes of characteristic size ~10-100 microns, where a regular delivery of flow tracers for µ-PIV may be difficult due to geometry, or they may clog microchannels.

The method can be used for a wide range of optically trappable micro-objects with spherical or non-spherical geometries, and smooth and rough surface morphology. Micropipette-driven flows allow rotating particles irrespectively to their optical anisotropy and trapping beam polarizations, being a simplified analog of optically actuated microtools.

We have experimentally demonstrated the ability to switch the kinetics of the very same microparticle between diffusion- and surface-controlled regimes, which unblocks a route towards precise single-particle dissolution studies and allows a controllable changing of a single particle size. Controllable dissociation of drug cargoes with the aid of external flows can be used as a continuous on-demand drug release trigger, which is one of the most important functions to grant future theragnosis particles.

## Acknowledgments

The research was supported by ERC StG "In Motion" (802279).

A.U. acknowledges the support of the Azrieli Foundation's Postdoctoral Fellowship.